\documentclass[pre,twocolumn,noshowpacs,showkeys,amsfonts,amssymb,amsmath,eqsecnum,letterpaper,nofootinbib,floatfix]{revtex4}

\usepackage{graphicx}
\begin{document}
\newlength{\GraphicsWidth}
\setlength{\GraphicsWidth}{8cm}     

\title{Non-linear screening of spherical and cylindrical colloids: the case 
of 1:2 and 2:1 electrolytes}

\author{Gabriel T\'ellez}
\email{gtellez@uniandes.edu.co}
\affiliation{Departamento de F\'{\i}sica, Universidad de Los Andes,
A.A.~4976, Bogot\'a, Colombia}
\author{Emmanuel Trizac}
\email{Emmanuel.Trizac@th.u-psud.fr}
\affiliation{Laboratoire de Physique Th\'eorique (Unit\'e Mixte de
Recherche UMR 8627 du CNRS), B\^atiment 210, Universit\'e de
Paris-Sud, 91405 Orsay Cedex, France}
 
\begin{abstract}
From a multiple scale analysis, we find an analytic solution 
of spherical and cylindrical Poisson-Boltzmann theory for 
both a 1:2 (monovalent co-ions, divalent counter-ions) 
and a 2:1 (reversed situation) electrolyte. Our approach
consists in an expansion in powers of rescaled 
curvature $1/(\kappa a)$, where $a$ is the colloidal radius 
and $1/\kappa$ the Debye length of the electrolytic solution. 
A systematic comparison with the full numerical solution
of the problem shows that for cylinders and spheres,
our results are accurate as soon as
$\kappa a>1$. We also report an unusual overshooting effect where the 
colloidal effective
charge is larger than the bare one.
\end{abstract}
 

\maketitle

\section{Introduction}
Almost a century ago, the work of Gouy \cite{Gouy} followed by that of
Chapman \cite{Chapman}, has established the foundations of the
mean-field treatment of the electric-double layer (Poisson-Boltzmann
theory).  This approach served as a basis for computing microionic
correlations in a homogeneous electrolyte \cite{DH} and later led to
the prominent DLVO theory of colloidal interactions \cite{DLVO}. An
essential notion in this context is that of charge renormalization
\cite{Belloni,Hansen,Levin,Trizac1,BocTrizAub-JCP}: at large
distances, the electrostatic signature of a charged body (with charge
$Z_\text{bare}$) in an electrolyte takes the same form as that of an
effective macro-ion with a suitable effective charge $Z_\text{eff}$,
the latter object being treated within linearized Poisson-Boltzmann
theory. Only for small $Z_\text{bare}$ do effective and bare
parameters coincide (weak coupling limit).  In
general, one has $|Z_\text{eff}| \ll
|Z_\text{bare}|$ which reflects the non-linear screening effect of the
electric double layer around a colloid \cite{rrq}. 
This non-linear regime, beyond
the weak coupling limit but below the couplings that would invalidate
the mean-field assumption underlying the approach, is precisely that
which is relevant for colloids (see e.g.~the discussion in references
\cite{Levin,Sat}).

Recently, analytical expressions have been obtained, within
Poisson-Boltzmann theory, for the effective charges of spherical and
cylindrical macro-ions \cite{ATB}. These predictions for a unique
macro-ion immersed in an infinite sea of monovalent electrolyte with
inverse Debye length $\kappa$, are exact up to $(\kappa a )^{-1}$
corrections, where $a$ is the radius of the macro-ion. For practical
purposes, the predictions are accurate as soon as $\kappa a >1$.  In
this paper, we consider the situation of spherical and cylindrical
macro-ions in a charge asymmetric electrolyte with both monovalent and
divalent micro-ions. The asymmetry of electrolyte has noticeable
consequences on the structure of the electric double-layer and the
case of 2:1 electrolytes (i.e.~with divalent co-ions and monovalent
counter-ions) turns out to differ much from the 1:2 situation
(monovalent co-ion, divalent counter-ion). Our analytical results
--obtained from a multiple scale technique \cite{Shkel}-- neglect
$O(\kappa a)^{-2}$ corrections for the electrostatic potential and
conversely, $O(\kappa a)^{-1}$ terms for effective charges.  By an
explicit comparison with the numerical mean-field results, they will
be shown to be precise whenever $\kappa a >1$, as was the case in
\cite{ATB}.  In section \ref{sec:pot}, the general method will be
presented, and the electrostatic potential obtained. The results
concerning effective quantities will be given in sections
\ref{sec:zeff} and \ref{sec:phieff}.  Conclusions will be drawned in
section \ref{sec:concl}.

\section{Quasi-planar solution to Poisson--Boltzmann equation for 2:1
or 1:2 electrolytes}
\label{sec:pot}

\subsection{2:1 electrolyte}

We consider a cylindrical ($j=1$) or spherical ($j=2$) colloid of
radius $a$ with surface charge density $e \sigma>0$ immersed in an
electrolyte with co-ions (resp.~counterions) of valency $z_1$
(resp.~$z_2$) and numeric density $n_1$ (resp.~$n_2$).  Let us
analyze in some detail the case $z_1=2$, $z_2=-1$, hereafter referred
to as 2:1.

As usual, we define the Debye length $\kappa^{-1}=(4\pi l_B \sum_i n_i
z_i^2)^{-1/2}= (12\pi n_2 l_B)^{1/2}$, the reduced electrical
potential $y=\beta e \psi$ and $\sigma^*=4\pi l_B \sigma a$. 
Here, $l_B$ denotes the Bjerrum length, defined from the
permittivity $\chi$ of the suspending medium and the inverse temperature
$\beta=1/(kT)$ as $l_B=\beta e^2/\chi$.
Using the
method of multiple scales, closely following~\cite{Shkel}, the
Poisson--Boltzmann equation
\begin{equation}
\label{eq:PB21}
\frac{1}{r^j}\frac{d}{dr}\left[r^j \frac{dy}{dr}\right]=
- 4\pi l_B n_2 (e^{-2y}-e^y)
\end{equation}
can be cast into
\begin{eqnarray}
\label{eq:PB21-multiple-scales}
\frac{\partial^2 y}{\partial x_1^2}+
\frac{2\epsilon\partial^2 y}{\partial x_1 \partial x_2}+
\frac{\epsilon j}{x_2}\frac{\partial y}{\partial x_1}
+\epsilon^2 \frac{\partial^2 y}{\partial x_2^2}
+ \frac{\epsilon^2 j}{x_2}\frac{\partial y}{\partial x_2}
=&&
\nonumber\\
-\frac{1}{3} (e^{-2y}-e^{y})
\end{eqnarray}
with boundary conditions
\begin{subequations}
\begin{eqnarray}
\label{eq:BC-contact}
\left[\frac{\partial y}{\epsilon \partial x_1}
+\frac{\partial y}{\partial x_2}\right]_{x_1=0,x_2=1}&=&-\sigma^*
\\
\label{eq:BC-infinity}
\lim_{x_1\to\infty, x_2\to\infty}
x^{j}_2\left[\frac{\partial y}{\epsilon \partial x_1}
+\frac{\partial y}{\partial x_2}\right]
&=&0.
\end{eqnarray}
\end{subequations}
Here, we have defined $\epsilon=(\kappa a)^{-1}$, $x_1=\kappa(r-a)$
and $x_2=r/a$. We seek for a solution as an expansion in powers of
$\epsilon$ which is supposed to be a small parameter: $y=y_0+\epsilon
y_1 +\cdots$.

The equation for the zeroth order term is Poisson--Boltzmann equation
for a planar interface
\begin{equation}
\label{eq:PB21-equation-ordre0}
\frac{\partial^2 y_0}{\partial x_1^2}=
-\frac{1}{3} (e^{-2y_0}-e^{y_0})
\end{equation}
which has been solved by Gouy in his pioneering work \cite{Gouy}
(see also Grahame~\cite{Grahame}). The solution reads
\begin{equation}
\label{eq:PB21-solut-ordre0}
y_0(x_1,x_2)=\ln\left(1+\frac{6 q}{(1-q)^2}\right)
\end{equation}
with the short-hand notation $ q= t(x_2) e^{-x_1} $, which will be
used extensively in the following. Here $t(x_2)$ is a function of
$x_2$ which appears as a constant of integration (with respect to
$x_1$) since in Eq.~(\ref{eq:PB21-equation-ordre0}) the variable $x_2$
does not appear. As explained in~\cite{Shkel} this function is
determined by the requirement that the non-homogeneous part of the
differential equation for the next order, $y_1$, decays faster than
$e^{-x_1}$ when $x_1\to\infty$. The equation for $y_1$ reads
\begin{equation}
\label{eq:PB21-equation-ordre1}
\frac{\partial^2 y_1}{\partial x_1^2}
-
\frac{1}{3}(2e^{-2y_0}+e^{y_0}) y_1
=
-\frac{2\partial^2 y_0}{\partial x_1 \partial x_2}
-\frac{j}{x_2}\frac{\partial y_0}{\partial x_1}
\end{equation}
The requirement that the r.h.s.~of Eq.~(\ref{eq:PB21-equation-ordre1})
decays faster than $e^{-x_1}$ leads to $t(x_2)=A x_2^{-j/2}$ with $A$ a
constant of integration. We therefore have
\begin{equation}
\label{eq:q}
q=A x_2^{-j/2} e^{-x_1} .
\end{equation}
Notice that the situation is exactly the same as in the 1:1
electrolyte case~\cite{Shkel}, the zero order solution in the
quasi-planar approximation is obtained from the planar solution with
the replacement of the constant of integration $A$ by $A
x_2^{-j/2}$. Actually, this is a general result for any type of
electrolyte since the r.h.s. of Eq.~(\ref{eq:PB21-equation-ordre1})
does not depend on the microscopic constitution of the electrolyte and
when $x_1\to\infty$ for any electrolyte the behavior of $y_0$ will be
given by the Debye-H\"uckel solution: $\text{cst} \times t(x_2)
\exp(-x_1)$.

The constant of integration $A$ can be expressed as a function of the
surface charge density $\sigma^*$ by enforcing the boundary
condition~(\ref{eq:BC-contact}) at the dominant order 
\begin{equation}
\left.\frac{\partial y_0}{\partial x_1}\right|_{x_1=0,x_2=1} = -s
\end{equation}
where we have put $s=\epsilon \sigma^*$.
This gives a third order equation for $A$
\begin{equation}
\label{eq:PB21-pour-A}
\frac{6A(1+A)}{(1-A)(A^2+4A+1)}=s.
\end{equation}
Its physical solution (which vanishes when $s\to0$) can be
written as
\begin{equation}
\label{eq:PB21-grand-A}
A=\frac{1}{s}
\left[
-2-s+2^{3/2}(2+s+s^2)^{1/2}\cos\left(\frac{\theta}{3}\right)
\right]
\end{equation}
with
\begin{equation}
\theta=\cos^{-1}\left[
\frac{-4-3s-3s^2-s^3}{\sqrt{2}(2+s+s^2)^{3/2}}
\right].
\end{equation}
This constant has also been computed in the study of the planar
interface effective charge ($6A=4\pi s_{\text{eff}}$) in
Ref.~\cite{Ulander}, although it is presented there in a slightly (but
completely equivalent) form.

Replacing the explicit expression~(\ref{eq:PB21-solut-ordre0}) for
$y_0$ into Eq.~(\ref{eq:PB21-equation-ordre1}) gives for the order-one
term $y_1$ the following equation
\begin{eqnarray}
\frac{\partial^2 y_1}{\partial x_1^2}
-
\frac{1+27 q^2+16q^3+27q^4+q^6}{(1-q)^2(1+4q+q^2)^2}
\,y_1
=&&
\nonumber\\
-\frac{12 j}{x_2}
\frac{q^2 (q^3+3q^2+3q-1)}{(1-q)^2(1+4q+q^2)^2}.
\end{eqnarray}
Using the variable $q$ instead of $x_1$ and performing the change of
function $y_1(x_1,x_2)= f(q)/[(1-q)(1+4q+q^2)]$ yields a second order
linear differential non-homogeneous equation for $f(q)$ with
polynomial coefficients in $q$ whose associated linear homogeneous
equation has the simple solution $f(q)=q(q+1)$, therefore
allowing to find the complete solution to the non-homogeneous equation
using the traditional method of ``variation of the constant''. After some
tedious but otherwise straightforward calculations, we find the
solution satisfying the appropriate boundary
condition~(\ref{eq:BC-infinity}) at infinity,
\begin{equation}
\label{eq:PB21-solut-partielle-ordre1}
y_1(x_1,x_2)= \frac{k(x_2) q (q+1) - \frac{j}{2x_2}
q^2(q^2+9q-8)}{(1-q)(1+4q+q^2)}.
\end{equation}
Again there is a function $k(x_2)$ that appears as a constant of
integration with respect to $x_1$ since there are no derivatives of
$y_1$ with respect to $x_2$ in Eq.~(\ref{eq:PB21-equation-ordre1}).
This function $k(x_2)$ is determined~\cite{Shkel} by the requirement
that the non-homogeneous part of the equation for the next order term
$y_2$ decreases faster that $e^{-x_1}$ when $x_1\to\infty$
\begin{eqnarray}
\frac{\partial^2 y_0}{\partial x_2^2}+\frac{j}{x_2}\frac{\partial
y_0}{\partial x_2} + 2 \frac{\partial^2 y_1}{\partial x_1\partial x_2}
+
\frac{j}{x_2}\frac{\partial y_1}{\partial x_1}
\\
+y_1^2 (e^{y_0}-4 e^{-2y_0})
&=&o(e^{-x_1})
\nonumber
\end{eqnarray}
This gives $k(x_2)=c_1+3j(j-2)/(4 x_2)$ with $c_1$ another constant of
integration, so finally the order-one solution is
\begin{equation}
\label{eq:PB21-solution-ordre1}
y_1(x_1,x_2)= \frac{ \left[c_1+\frac{3j(j-2)}{4 x_2}\right] 
q (q+1) - \frac{j}{2x_2}
q^2(q^2+9q-8)}{(1-q)(1+4q+q^2)}
\end{equation}
Applying the boundary condition~(\ref{eq:BC-contact}) to the next order
in $\epsilon$ gives the equation $\left.\partial_{x_1}
y_1+\partial_{x_2} y_0 \right|_{x_1=0,x_2=1}=0$ and
subsequently determines the
constant of integration 
\begin{widetext}
\begin{equation}
\label{eq:PB21-petit-c1}
c_1=-j
\frac{
2A^6+12 A^5+
A^4(34+3j)
+2A^3(-88+3j)+6 A^2(-7+3j) 
+2A(34+3j)
+3(2+j)
}{4(1+2A+6A^2+2A^3+A^4)}.
\end{equation}
\end{widetext}
The quantity $A$ is given by Eq.~(\ref{eq:PB21-grand-A}). Both constants $A$ and
$c_1$ are related to the effective charge of the colloid and
therefore carry important physical information about the
system. Let us notice that at saturation $\sigma\to\infty$, they take
simple values: $A^{\text{sat}}=1$ and
$c_1^{\text{sat}}=-j(3j-8)/4$.

\subsection{1:2 electrolyte}

The quasi-planar approximate solution of the Poisson--Boltzmann
equation for the case $z_1=1$ and $z_2=-2$ (1:2 electrolyte) follows
from similar calculations. We only report the results. The zero order
term $y_0$ reads
\begin{equation}
\label{eq:PB12-solut-ordre0}
y_0(x_1,x_2)=
-\ln\left(
1-\frac{6q}{(q+1)^2}
\right)
\end{equation}
with $q$ given by Eq.~(\ref{eq:q}) and the order-one term is
\begin{equation}
\label{eq:PB12-solut-ordre1}
y_1(x_1,x_2)=
\frac{ -q(q-1)\left[c_1+\frac{3j(j-2)}{4 x_2}\right] 
 + \frac{j}{2x_2}
q^2(q^2-9q-8)}{(1+q)(1-4q+q^2)}
\end{equation}
Note that the solution for the 1:2 case is simply obtained from the
one for the 2:1 case by a global change of sign and by replacing $q$
by $-q$.

The constant of integration $A$ is again a solution of a third order
equation which can be obtained from Eq.~(\ref{eq:PB21-pour-A}) by a
global change of sign and by replacing $A$ by $-A$. However, the
physical solution is not the same as in the 2:1 case, and now takes the form
\begin{equation}
\label{eq:PB12-grand-A}
A=\frac{1}{s}
\left[
-2+s+2^{3/2}(2-s+s^2)^{1/2}\cos\left(\frac{\theta+4\pi}{3}\right)
\right],
\end{equation}
with $\theta$ given by
\begin{equation}
\theta=\cos^{-1}\left[
\frac{-4+3s-3s^2+s^3}{\sqrt{2}(2-s+s^2)^{3/2}}
\right].
\end{equation}
The constant of integration for the order-one term is here 
\begin{widetext}
\begin{equation}
\label{eq:PB12-petit-c1}
c_1=-j
\frac{
2A^6-12 A^5+
A^4(34+3j)
-2A^3(-88+3j)
+6 A^2(-7+3j) 
-2A(34+3j)
+3(2+j)
}{4(1-2A+6A^2-2A^3+A^4)},
\end{equation}
\end{widetext}
with $A$ given by Eq.~(\ref{eq:PB12-grand-A}). The saturation
($s\to\infty$) values of these constants are now different. We have
$A^{\text{sat}}=2-\sqrt{3}$ and
$c_1^{\text{sat}}=-j(28+3j-24\sqrt{3})/4$.

\subsection{Comparison between analytical and numerical potential profiles}
\begin{figure}
\includegraphics[width=6cm,angle=0]{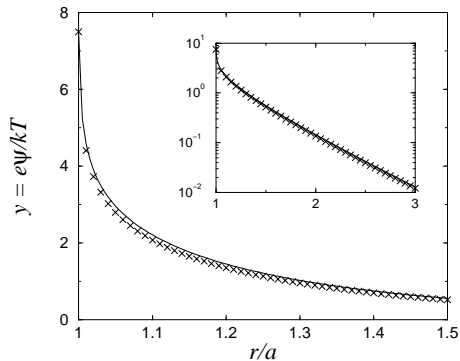}
\caption{Reduced electrostatic potential $y(r)$ as a function 
of rescaled distance for a spherical macro-ion in a 1:2 electrolyte. 
The continuous curve shows the numerical solution of the problem
and the crosses indicate the values found from Eq. 
(\ref{eq:PB12-solut-ordre0}) and (\ref{eq:PB12-solut-ordre1}).
The inset shows the same data on a linear-log scale. Here,
$\kappa a = 2$ and the reduced bare charge is very high:
$Z_{\text{bare}\,} l_B/a = 2000$.
\label{fig:pot12}
}
\end{figure}

\begin{figure}
\includegraphics[width=6cm,angle=0]{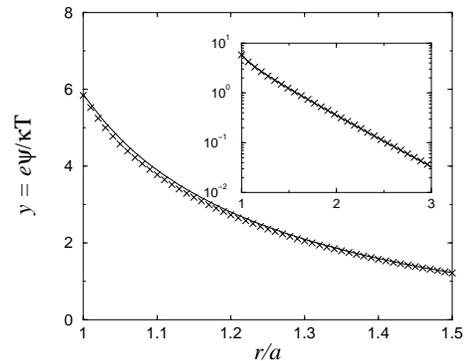}
\caption{Same as Fig. \ref{fig:pot12} in a 2:1 electrolyte. Here,
$\kappa a = 2$ and the reduced bare charge is:
$Z_{\text{bare}\,} l_B/a = 34$.
\label{fig:pot21}
}
\end{figure}

Gathering results, we obtain up to corrections of order $1/(\kappa a)^2$, 
$y(r)=y_0(r)+ (\kappa a)^{-1} y_1(r)$,
where the auxiliary functions $y_0$ and $y_1$ are given by 
Eqs. (\ref{eq:PB21-solut-ordre0}) and (\ref{eq:PB21-solution-ordre1}) in the 2:1 case, and by
Eqs. (\ref{eq:PB12-solut-ordre0}) and (\ref{eq:PB12-solut-ordre1}) 
for 1:2 electrolytes. It is instructive to compare the resulting predictions
to the numerical solution of Poisson-Boltzmann theory,
obtained following the method of Ref. \cite{Lang}.
Figures \ref{fig:pot12} and \ref{fig:pot21} show that already
for $\kappa a=2$, the agreement is good. Although the potential
at contact $y(a)$ is predicted accurately, we observe that our 
theoretical expression slightly underestimates the potential.
A similar trend will be observed for effective charges 
--again for spheres-- 
in section \ref{sec:zeff}. In cylindrical geometry, a slight overestimation
may be found in the 1:2 case.

The parameters in figures~\ref{fig:pot12} and~\ref{fig:pot21} are
chosen to be in the non-linear saturation regime $Z_{\text{bare}}\gg
a/l_B$. It is interesting to notice that the relative error of our
analytic solution from the numerical one in the cases presented in
figures~\ref{fig:pot12} and~\ref{fig:pot21} is of order 3\% that is of
order $(\kappa a)^{-2}/10$. We have also studied the linear regime,
$Z_{\text{bare}}$ small, and in this case the error is larger of order
25\%, i.e.~of order $(\kappa a)^{-2}$ (remember that in our analytical
solution we neglect terms of order $(\kappa a)^{-2}$). We have also
computed the relative error for other values of $\kappa a$ and the
trend is general: in the linear regime the relative error is of
order $(\kappa a)^{-2}$ but for the non-linear saturation regime the
situation improves and the error is reduced by a factor 10. This makes
our analytic solution practical since experimental
situations are often in the saturation regime where our solution is more
accurate.

\section{Effective charges}
\label{sec:zeff}

\subsection{Spheres}

\begin{figure}
\includegraphics[height=6cm]{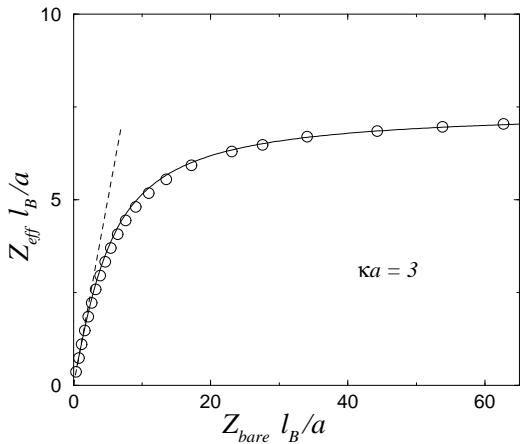}
\caption{Effective vs bare charge for a spherical macro-ion in a 1:2
electrolyte (i.e. monovalent co-ions/divalent counter-ions).
The open circles are obtained from the full non-linear Poisson Boltzmann
theory, while the continuous curve corresponds to the analytical prediction 
given by Eqs. (\ref{eq:Zeff}). The dashed line has slope 1 and shows the 
initial linear regime for weak charges. The salinity conditions are here
such that $\kappa a =3$ where $a$ is the sphere radius. 
\label{fig:Zeff12}
}
\end{figure}

\begin{figure}[hbt]
\includegraphics[height=6cm]{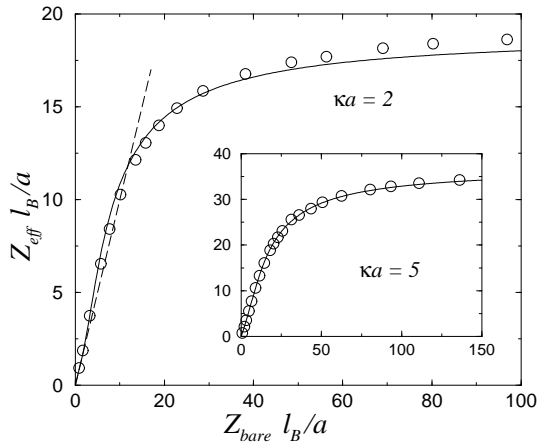}
\caption{Same as Figure \ref{fig:Zeff12} for a 2:1
electrolyte (divalent co-ions, monovalent counter-ions).
As indicated, the main graph corresponds to $\kappa a =2$
while the inset shows results for $\kappa a = 5$.
\label{fig:Zeff21}
}
\end{figure}

The far field $r\to\infty$ behavior of the solution
$y(r)=y_0(r)+\epsilon y_1(r)+O(\epsilon^2)$, obtained in the last section
is
\begin{equation}
\label{eq:far-field}
y(r)\underset{r\to\infty}{\sim} A e^{-\kappa(r-a)}
\left(\frac{a}{r}\right)^{j/2} \left( 6 +
\frac{c_1}{\kappa a}\right)+O(\epsilon^2).
\end{equation}
With this expression, we can deduce the effective charge. For 
a spherical macro-ion ($j=2$) of radius $a$ and charge 
$Z_\text{eff}$, the solution of linearized Poisson-Boltzmann
theory (also referred to as Debye-H\"uckel theory)
$\nabla^2 y=\kappa^2 y$ reads
\begin{equation}
\label{eq:far-field-DH}
y(r) =
\frac{Z_{\text{eff}}\,l_B}{1+\kappa a}
\frac{e^{-\kappa(r-a)}}{r}.
\end{equation}
By comparison with expression (\ref{eq:far-field})
we conclude that the effective charge is given by
\begin{equation}
Z_{\text{eff}} \frac{l_B}{a}
=
A\left[6\kappa a + 6 +c_1+ O\left(\frac{1}{\kappa a}\right) \right].
\label{eq:Zeff}
\end{equation}
The coefficients $A$ and
$c_1$ are given by Eqs.~(\ref{eq:PB21-grand-A}) and
(\ref{eq:PB21-petit-c1}) (2:1 electrolyte) or
Eqs.~(\ref{eq:PB12-grand-A}) and (\ref{eq:PB12-petit-c1}) (1:2
electrolyte) in terms of the bare charge $Z_{\text{bare}}$ by
reporting $s= \epsilon Z_{\text{bare}} l_B/a $.
Figures \ref{fig:Zeff12} and \ref{fig:Zeff21}
compare the above analytical predictions
to the effective charge obtained from the far field behavior
of the numerical solution of Poisson-Boltzmann theory,
obtained as explained in \cite{Lang}. The agreement
is satisfying, and improves upon increasing $\kappa a$, as was
anticipated.

One may readily check from Eq. (\ref{eq:Zeff}) 
that in the limit $Z_\text{bare}\to 0$, 
$Z_\text{eff}/Z_\text{bare} \to 1$. Effective and bare parameters
coincide in the weak coupling limit, 
as it should (see the dashed lines in Figures \ref{fig:Zeff12}
and \ref{fig:Zeff21}). 
In the other limit where
$Z_\text{bare} \to \infty$, we observe the saturation
picture common to several mean-field theories \cite{Trizac1,Sat}:
the effective charge goes to a plateau value, that only depends
on two dimensionless quantities, $a/l_B$ and $\kappa a$.
The effective charge at saturation (obtained when $s\to\infty$) takes a 
simple expression. For a 2:1 electrolyte 
\begin{equation}
\label{eq:PB21-Zsat}
Z_{\text{eff}}^{\text{sat}}
\frac{l_B}{a}
=
6\kappa a + 7 + O\left(\frac{1}{\kappa a}\right)
\end{equation}
and for a 1:2 electrolyte
\begin{eqnarray}
\label{eq:PB12-Zsat}
Z_{\text{eff}}^{\text{sat}}
\frac{l_B}{a}
&=&
\left(2-\sqrt{3}\right)
\left[6\kappa a - 11 + 12\sqrt{3} + 
O\left(\epsilon\right)
\right]
\\
&\simeq&
1.608\, \kappa a + 2.623 + 
O\left(\frac{1}{\kappa a}\right).
\nonumber
\end{eqnarray}
We note that the conditions of Fig. \ref{fig:pot12} are those
of saturation.

\begin{figure}[htb]
\includegraphics[height=6cm,angle=0]{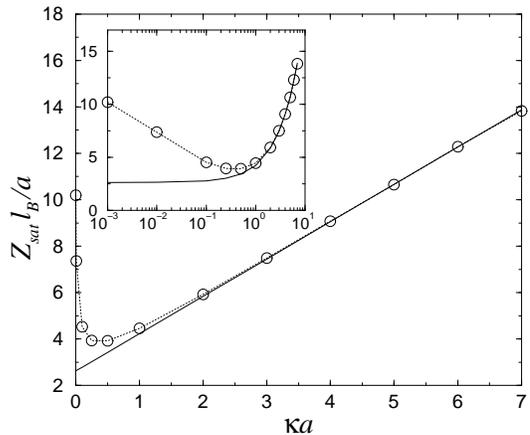}
\caption{Effective charge at saturation in a 1:2
electrolyte for a spherical colloid. 
The line shows the prediction of Eq. (\ref{eq:PB12-Zsat}).
The circles again correspond to the numerical resolution of Poisson-Boltzmann
theory, and the dotted line between them is a guide to the eye.
The inset shows the same data in a log-linear scale.
\label{fig:Zsat12}
}
\end{figure}

\begin{figure}[htb]
\includegraphics[height=6cm,angle=0]{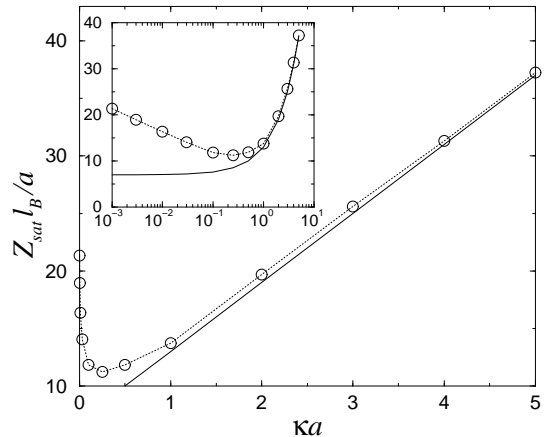}
\caption{Same as Figure \ref{fig:Zsat12} for a 2:1
electrolyte. The line shows the prediction of Eq. (\ref{eq:PB21-Zsat}).
\label{fig:Zsat21}
}
\end{figure}

These expressions are tested against the numerical data in Figures
\ref{fig:Zsat12} and \ref{fig:Zsat21}. The agreement is good for 
$\kappa a > 1$. In Figs.~\ref{fig:Zsat12} and \ref{fig:Zsat21},
an inset has been added
to show the regime of low $\kappa a $ values where our approach breaks.
In this limit, the observed divergence of $Z_{\text{eff}}^{\text{sat}}$ means 
that the bare charge is no longer renormalized.
As happens for monovalent electrolytes \cite{Belloni}, 
the saturated effective charge is a non-monotonous function of
$\kappa a$, that reaches its minimum for $\kappa a \simeq 0.3$.
In the latter 1:1 case, we recall for completeness that the
asymptotic expansion $Z_{\text{eff}}^{\text{sat}} l_B/a = 4 \kappa a +6$
holds for $\kappa a >1$ \cite{ATB}.

\subsection{Cylinders}

\begin{figure}[t]
\includegraphics[height=6cm,angle=0]{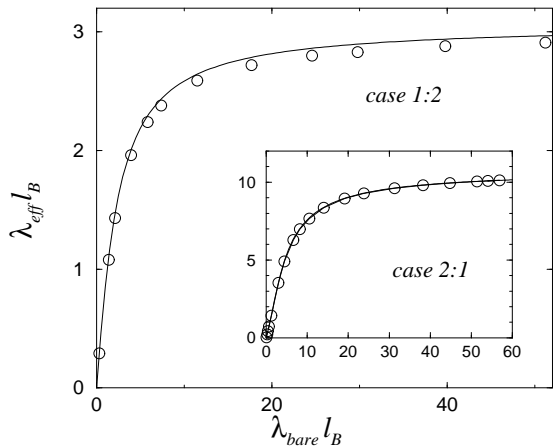}
\caption{Effective line charge density as a function of its bare
counterpart, for a rod-like macro-ion.  The line shows the prediction
of Eq. (\ref{eq:laeff}) and the symbols stand for the ``exact''
numerical values. The main graph and the inset correspond to the same
salinity conditions $\kappa a =3$, where $a$ is the cylinder's radius.
\label{fig:laeff12}
}
\end{figure}

For an infinite long cylindrical colloid, $j=1$, with linear charge
density $e \lambda$, the far-field solution~(\ref{eq:far-field}) should
be compared to the one obtained from Debye--H\"uckel theory,
\begin{eqnarray}
y(r)
&=&
\frac{2 l_B \lambda_{\text{eff}}}{\kappa a}
\frac{K_0(\kappa r)}{K_1(\kappa a)}
\\
&\underset{r\to\infty}{\sim}&
\frac{2 l_B \lambda_{\text{eff}}}{\kappa a}
\frac{\sqrt{\frac{\pi}{2\kappa a}} e^{-\kappa a}}{K_1(\kappa a)}
\left(\frac{a}{r}\right)^{1/2} e^{-\kappa(r-a)}.
\nonumber
\end{eqnarray}
where $K_0$ and $K_1$ are modified Bessel functions. We conclude that
the effective line charge density is given by
\begin{equation}
\lambda_{\text{eff}} l_B
= A \left[3 \kappa a +
\frac{9}{8}+\frac{c_1}{2} +O\left(\frac{1}{\kappa a}\right)
\right].
\label{eq:laeff}
\end{equation}
The explicit expression of the effective linear charge density in
terms of the bare linear charge density $\lambda_{\text{bare}}$ is
obtained by reporting $s=2 \epsilon l_B \lambda_{\text{bare}}$ in the
analytical expressions for $A$ and $c_1$ given in
Eqs.~(\ref{eq:PB21-grand-A}) and (\ref{eq:PB21-petit-c1}) (2:1
electrolyte) or Eqs.~(\ref{eq:PB12-grand-A}) and
(\ref{eq:PB12-petit-c1}) (1:2 electrolyte).
Figure \ref{fig:laeff12} shows the accuracy of our analytical expression,
that turns out to be slightly better in the 2:1 situation than in the 1:2 case
(the reverse observation follows from inspecting Figures 
\ref{fig:Zsat12} and \ref{fig:Zsat21}).

\begin{figure}[t]
\includegraphics[height=6cm,angle=0]{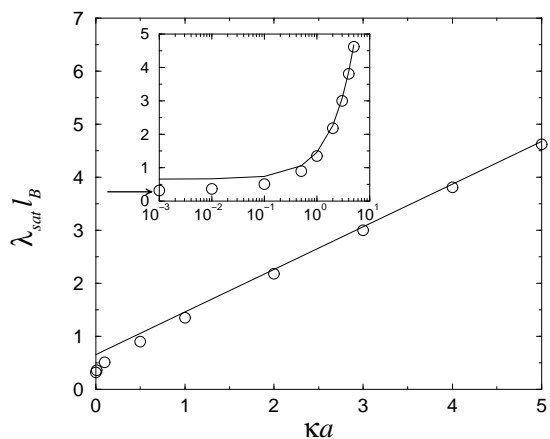}
\caption{Saturated effective line charge of an infinite cylinder 
with radius $a$ in a 1:2 electrolyte. Line: Eq. (\ref{eq:lasat12}) and symbols:
numerical solution. The inset is a log-linear plot. The
arrow indicates the value $\sqrt{3}/(2\pi)\simeq 0.275$, 
obtained in the $\kappa a \to 0$ limit \cite{Tracy,prep}.
\label{fig:lasat12}
}
\end{figure}

\begin{figure}[t]
\includegraphics[height=6cm,angle=0]{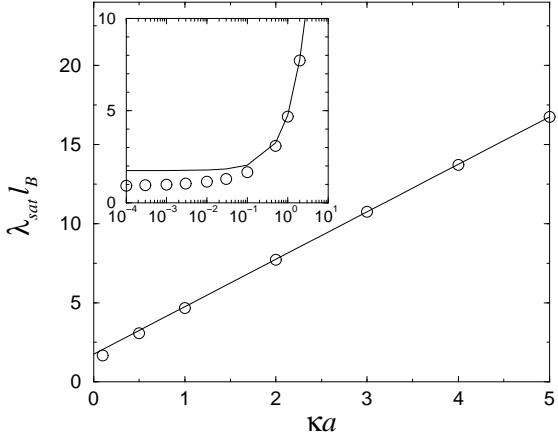}
\caption{Same as Figure \ref{fig:lasat12} for a 2:1 electrolyte.
The line corresponds to Eq. (\ref{eq:lasat21})
\label{fig:lasat21}
}
\end{figure}

The effective charges at saturation are, for a 2:1 electrolyte,
\begin{equation}
\lambda_{\text{eff}}^{\text{sat}}
 l_B
=  3 \kappa a +
\frac{7}{4} +
O\left(\frac{1}{\kappa a}\right)
\label{eq:lasat21}
\end{equation}
and for a 1:2 electrolyte,
\begin{eqnarray}
\label{eq:lasat12}
\lambda_{\text{eff}}^{\text{sat}} l_B &=& 
\left(2-\sqrt{3}\right)
 \left[3 \kappa a
-\frac{11}{4}+3\sqrt{3} 
+
O\left(\epsilon\right)
\right]
\\
&\simeq&
0.804\, \kappa a + 0.655 
+
O\left(\frac{1}{\kappa a}\right).
\nonumber
\end{eqnarray}
These simple expressions are plotted in Figures \ref{fig:lasat12}
and \ref{fig:lasat21}, together with their counterparts
obtained from the solution of Poisson-Boltzmann theory,
shown by symbols. When converted into effective surface charge
densities $\sigma_\text{eff}^{\text{sat}}$, the previous results yield,
up to $(\kappa a)^{-1}$ corrections
\begin{eqnarray}
4 \pi a l_B \sigma_\text{eff}^{\text{sat}} &=&  6 \kappa a + 7 
\quad (2:1, \hbox{ spheres}) \label{eq:311}\\
&=& 6 \kappa a + \frac{7}{2} 
\quad (2:1, \hbox{ rods}),\label{eq:312}
\end{eqnarray}
while in the planar case, one gets 
$4 \pi a l_B \sigma_\text{eff}^{\text{sat}} =  6 \kappa a +0$.
The increase of the zeroth order term ($0,7/2,7$) as the
dimensionality of the object increases, reflects the concomitant
weaker range of the bare Coulomb potential ($-r$ in 1D, $-\log r$ in
2D, $1/r$ in 3D\ldots). Indeed, a weaker Coulomb contribution 
leads to a weaker screening, hence a higher effective charge.
A similar argument therefore explains the increase of effective 
charges with $\kappa a$. For completeness, we also give the 1:2 results
\begin{eqnarray}
\!\!\!\!\!
\frac{4 \pi a l_B \sigma_\text{eff}^{\text{sat}}}{\left(2-\sqrt{3}\right)} &=&  
\left[6\kappa a - 11 + 12\sqrt{3}  \right]
\quad (\hbox{spheres})~ \\
&=&  \left[6 \kappa a-\frac{11}{2}+6\sqrt{3} \right]
\quad (\hbox{rods}),~
\end{eqnarray}
and the same argument as above equally applies here.

It may be observed in Figures \ref{fig:lasat12} and \ref{fig:lasat21}
that in cylindrical geometry, the saturated effective charge does not
diverge in the limit $\kappa a \to 0$, as was the case for spheres due
to entropic reasons (Boltzmann ``beats'' Coulomb in this limit).  With
infinite cylinders, this is no longer the case (a two dimensional
situation is more favorable to Coulomb than a 3D one), and
$\lambda_{\text{eff}}^{\text{sat}}$ reaches a constant value for small
$\kappa a$ (see the inset of Figures \ref{fig:lasat12}
and~\ref{fig:lasat21}).

\subsection{An overshooting effect}
\begin{figure}[t]
\includegraphics[height=6cm,angle=0]{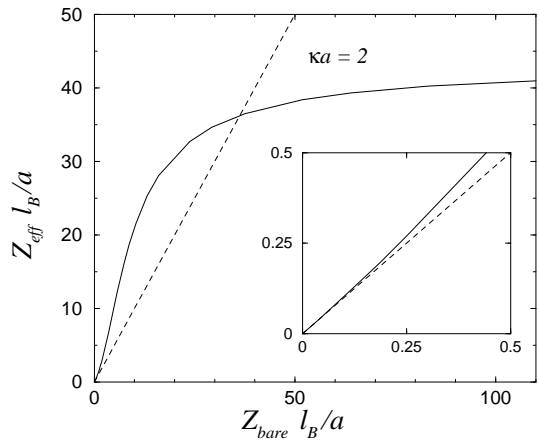}
\caption{Illustration of the overshooting effect, for a spherical 
colloid in a 5:1 electrolyte, with $\kappa a=2$. 
The dashed line has slope 1 and
the inset (a zoom on the bottom left corner) shows that for 
small charges, $Z_\text{eff}$ is a convex-up function of 
$Z_\text{bare}$.
\label{fig:overshoot}
}
\end{figure}

Although the effect is not very marked, it may be observed in
Fig.~\ref{fig:Zeff21} that $Z_\text{eff}$ as a function
$Z_\text{bare}$ has an inflexion point, so that $Z_\text{eff} >
Z_\text{bare}$ in a given charge range ($0<Z_\text{eff}<10 a/l_B$,
where not only our prediction but also the symbols showing numerical
data lie above the dashed line).  This was unexpected since with a
monovalent (1:1) electrolyte, the effective charge is always smaller
than the bare one.  This overshooting effect occurs for spherical but
also for rod-like macroions. It requires a 2:1 salt for which the
divalent co-ions are expelled from the vicinity of the macro-ion,
which leads to a much weaker screening than in the reverse 1:2
situation. That this effect is able to impose $Z_\text{eff} >
Z_\text{bare}$ is however surprising, and in order to check its
robustness, we also investigated numerically more asymmetric
electrolytes. Figure \ref{fig:overshoot} shows that for a 5:1 salt,
the overshooting is more pronounced and that for $Z_\text{eff} \simeq
10 a/l_B$, the effective charge may be twice bigger than the bare one.

Additional insight into this unexpected overshooting effect may be
obtained from our analytical expressions for the effective charges
Eqs.~(\ref{eq:Zeff}) and~(\ref{eq:laeff}). For a small surface charge
density $s$, in the 2:1 case, the quantities $A$ and $c_1$ involved in
the expressions of the effective charges have a Taylor expansion of
the form
\begin{subequations}
\label{eq:A-c-small-s-2:1}
\begin{eqnarray}
  A&=&\frac{s}{6}+\frac{s^2}{18}-\frac{s^3}{216}+O(s^4)\\
  c_1&=&-\frac{9}{4}-\frac{7s}{3}+\frac{13s^2}{24}+\frac{10s^3}{9}+
  O(s^4);\ \text{rods}\\
  c_1&=&-6-\frac{14s}{3}+\frac{13s^2}{12}+\frac{20s^3}{9}+O(s^4);
  \ \text{spheres}
  \nonumber\\
\end{eqnarray}
\end{subequations}
For rods, this gives the following behavior of the effective linear
charge density for small bare charge
\begin{equation}
  \lambda_{\text{eff}}l_B=
  \lambda_{\text{bare}}l_B
  + \frac{s^2}{6}\left(\kappa a-\frac{7}{6}\right)
  -\frac{s^3}{72}\left(\kappa a+\frac{17}{12}\right)
  +O(s^4)
\end{equation}
In this case $s=2\lambda_{\text{bare}}l_B(\kappa a)^{-1}$. For
spherical colloids, the behavior for small bare
charges is
\begin{equation}
  Z_{\text{eff}}\frac{l_B}{a}=
  Z_{\text{bare}}\frac{l_B}{a}
  + \frac{3s^2}{9}\left(\kappa a-\frac{7}{3}\right)
  -\frac{s^3}{36}\left(\kappa a+\frac{17}{6}\right)
  +O(s^4)
\end{equation}
with now $s=Z_{\text{bare}}l_B/(\kappa a^2)$. Remembering that our
analytical solution is valid for large values of $\kappa a$, we notice
that in both cases the coefficient of the term of order two in the
bare surface charge density $s$ is positive. This implies that the
effective charge will become larger than the bare charge in a certain
intermediate regime of values of the bare charge when non-linear
effects start to become important (being nevertheless far form the
strongly non-linear saturated regime where the effective charge
saturates). Let us mention that
this overshooting effect is also expected for a planar geometry, since
in that case the effective charge is essentially given by $A$.

In contrast with this, in the case of a 1:2 electrolyte the Taylor
expansions of $A$ and $c_1$ are similar to
Eqs.~(\ref{eq:A-c-small-s-2:1}) with the formal replacement
$s\rightarrow -s$ and $A\rightarrow -A$, in particular the sign of the
order $s^2$ in $A$ changes sign. This gives the following behavior
for the effectives charges
\begin{equation}
  \lambda_{\text{eff}}l_B=
  \lambda_{\text{bare}}l_B
  - \frac{s^2}{6}\left(\kappa a-\frac{7}{6}\right)
  -\frac{s^3}{72}\left(\kappa a+\frac{17}{12}\right)
  +O(s^4)
\end{equation}
for rods and 
\begin{equation}
  Z_{\text{eff}}\frac{l_B}{a}=
  Z_{\text{bare}}\frac{l_B}{a}
  - \frac{3s^2}{9}\left(\kappa a-\frac{7}{3}\right)
  -\frac{s^3}{36}\left(\kappa a+\frac{17}{6}\right)
  +O(s^4)
\end{equation}
for spheres. The coefficient of $s^2$ has changed sign with respect
to the 2:1 case. This coefficient is now negative and this implies
that the effective charge will remain smaller that the bare charge,
thus no overshooting effect for the 1:2 electrolyte case.

It is interesting to mention that for a symmetric 1:1 electrolyte, the
small bare charge behavior reads~\cite{ATB}
\begin{equation}
  \lambda_{\text{eff}}l_B=
  \lambda_{\text{bare}}l_B
  -\frac{x_r^3}{4}\left(\kappa a - \frac{1}{4}\right)+
  O(x_r^5)
\end{equation}
for rods with $x_r=\lambda_{\text{bare}}l_B/[\kappa a+(1/2)]$ and
\begin{equation}
  Z_{\text{eff}}\frac{l_B}{a}=
  Z_{\text{bare}}\frac{l_B}{a}
  -\frac{x_s^3}{2}\left(\kappa a-\frac{1}{2}\right)+O(x_s^5)
\end{equation}
for spheres with $x_s=Z_{\text{bare}}l_B/[2a(\kappa
a+1)]$. Interestingly, there is no term of order two in the bare
charge as opposed to the asymmetric electrolytes cases. The first
correction to the linear term is of order three and it is negative.
The effective charge will be smaller than the bare charge: no
overshooting effect here either.

The first corrections to the linear theory are of order two in the
bare charge for asymmetric electrolytes and with the sign of
$\sum_\alpha z_\alpha^3 n_\alpha$, $z_\alpha$ being the valency of
species $\alpha$ and
$n_\alpha$ its density. On the other hand, 
for symmetric electrolytes, the first
correction is of order three in the bare charge. This important
difference between symmetric and asymmetric electrolytes also appears
in others contexts, namely in the study of the contributions due to
correlations to the effective charge, in a framework going beyond
the mean field approximation~\cite{AquaCornu1, AquaCornu2,
Tellez-3Dslab}.

Finally, we mention that recent HNC integral equation computations
for the same systems as investigated here 
confirm the validity of the overshooting
effect \cite{DL}. 
Explicit comparisons with our predictions are under way \cite{DL}. 

\section{Colloids as constant potential objects}
\label{sec:phieff}

Colloids are usually highly charged so that their effective charge 
--within mean-field--
is saturated and therefore independent of the bare one. Yet, the bare
charge is often not large enough to meet the region of high micro-ionic
electrostatic correlations where the mean field approach would break down
\cite{Levin,Sat}. This remark has led to the proposal to consider
highly charged colloids as objects of fixed effective potential in the
case of a 1:1 electrolyte \cite{Trizac1}. Similar considerations
may be put forward here. From the analysis of section 
\ref{sec:zeff}, the surface potentials $y=e \psi /(kT)$ associated with 
effective charges read, for spheres
\begin{eqnarray}
y_\text{eff}^{\text{sat}} &=&  
6 + \frac{1}{1+\kappa a}; \quad (2:1) \label{eq:poteffa}\\
y_\text{eff}^{\text{sat}}
&=& 6 (2-\sqrt{3}) + (2-\sqrt{3}) 
\frac{12 \sqrt{3}-17}{1 + \kappa a};
\quad (1:2)\nonumber
\end{eqnarray}
The important point is that the $\kappa a$ dependence is very weak
for $\kappa a>1$, which reinforces the picture of constant potential
objects. One may therefore consider a highly charged sphere as an effective
body of potential $6 kT/e$ or $6 (2-\sqrt{3}) kT/e$ depending on
2:1 or 1:2 asymmetry, irrespective of physico-chemical parameters.
In a 1:1 salt, one gets a value $4kT/e$ \cite{Trizac1}. It is 
natural to find this quantity in between the two bounds 
$6 kT/e$ and $6 (2-\sqrt{3}) kT/e$, since screening is all the more efficient
as the valency of counter-ions is large and the valency of co-ions is low
(in absolute values).

\begin{figure}[t]
\includegraphics[height=6cm,angle=0]{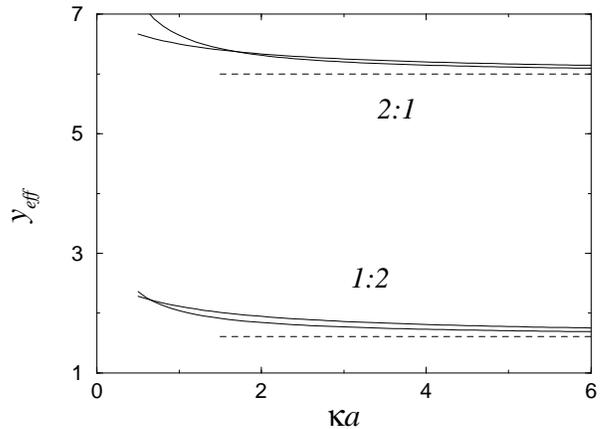}
\caption{Effective potentials at saturation following
from Eqs. (\ref{eq:poteffa}) and (\ref{eq:poteffb}), 
as a function of $\kappa a$. The limiting values
for $\kappa a \to \infty$ are shown by the dotted lines.
\label{fig:poteff}
}
\end{figure}

For rod-like polyions, we  get 
\begin{eqnarray}
y_\text{eff}^{\text{sat}} &=&  
\left(6 + \frac{7}{2\kappa a}\right)
\frac{K_0(\kappa a)}{K_1(\kappa a)}; \quad (2:1) 
\label{eq:poteffb}\\
y_\text{eff}^{\text{sat}}
&=& 6 (2-\sqrt{3}) + (2-\sqrt{3})\,
\frac{6 \sqrt{3}-11/2}{\kappa a}\frac{K_0(\kappa a)}{K_1(\kappa a)};
\quad (1:2)\nonumber
\end{eqnarray}
Expressions (\ref{eq:poteffa}) and (\ref{eq:poteffb})
are plotted in Fig. \ref{fig:poteff}.

\section{Conclusion}
\label{sec:concl}

In conclusion, we have found an analytical
solution of cylindrical and spherical Poisson-Boltzmann 
equation in asymmetric 1:2 and 2:1 electrolytes. Our approach
amounts to performing a curvature expansion, and neglects corrections
of order $1/(\kappa a)^2$ for the electrostatic potential.
For $\kappa a >1$, the corresponding solution and associated 
effective charge are in excellent agreement with their
counterparts obtained from the full numerical resolution of the
problem.

Our multiple scale analysis relies on the possibility to solve
analytically the planar problem (corresponding to $\kappa a \to
\infty)$. Since for a $n$:$m$ electrolyte (where $n$ and $m$
respectively stand for the valency of co-ions and counter-ions), this
solution is only known explicitly for $n/m=1,1/2$ and 2, we focused on
1:2 and 2:1 salts. The monovalent 1:1 situation has been investigated
in \cite{Shkel,ATB}. For a given salinity $\kappa$, non-linear
screening is more efficient with divalent than with monovalent
counter-ions.  Accordingly, we always found higher effective charges
in the 2:1 than in the 1:2 situation. Surprisingly, we found that 2:1
screening is even able to drive the effective charge in a regime where
it is higher than the bare one. This overshooting effect happens in an
intermediate charge range, since when $Z_\text{bare}$ is large enough,
$Z_\text{eff}$ saturates.


\begin{acknowledgments}
This work was supported by a ECOS Nord/COLCIENCIAS-ICETEX-ICFES action
of French and Colombian cooperation. G.~T. acknowledge partial
financial support from COLCIENCIAS under project 1204-05-13625.
\end{acknowledgments}



\begin{thebibliography}{99}
\bibitem{rrq}
We will however report here some situations where
$|Z_\text{eff}| > |Z_\text{bare}|$.
\end{thebibliography}
\end{document}